\documentclass[iop]{emulateapj}

\newcommand{\cjaa}{Chin. J. Astron. Astrophys.}

\slugcomment{}

\shorttitle{Early afterglow of gravitational wave bursts}
\shortauthors{Zhang}

\begin{document}


\title{Early X-ray and optical afterglow of gravitational wave bursts 
from mergers of binary neutron stars}


\author{Bing Zhang\altaffilmark{1,2,3}}


\altaffiltext{1}{Kavli Institute of Astronomy and Astrophysics, 
Peking University, Beijing 100871, China}
\altaffiltext{2}{Department of Astronomy, Peking University, 
Beijing 100871, China}
\altaffiltext{3}{Department of Physics and Astronomy, University of 
Nevada Las Vegas, NV 89154, USA}


\begin{abstract}
Double neutron star mergers are strong sources of gravitational waves. The
upcoming advanced gravitational wave detectors  
are expected to make the first detection of gravitational 
wave bursts (GWBs) associated with these sources. Proposed electromagnetic
counterparts of a GWB include a short GRB, an
optical macronova, and a long-lasting radio afterglow. Here we
suggest that at least some GWBs could be followed by an early afterglow
lasting for thousands of seconds, if the post-merger product is a
highly magnetized, rapidly rotating, massive neutron star rather 
than a black hole. This afterglow
is powered by dissipation of a proto-magnetar wind. The X-ray flux is
estimated to be as bright as $(10^{-8}-10^{-7})~{\rm erg~s^{-1}~cm^{-2}}$. The
optical flux is subject to large uncertainties but could be as bright as
17th magnitude in R-band. We provide observational hints of such a 
scenario, and discuss the challenge and strategy
to detect these signals.
\end{abstract}


\keywords{}



\section{Introduction}

Mergers of neutron-star/neutron-star (NS-NS) binaries are strong sources of
gravitational waves \citep[e.g.][]{kramer06}. The upcoming advanced
gravitational wave detectors such as Advanced LIGO \citep{ligo} and
Advanced VIRGO \citep{virgo} are expected to expand the detection
horizon to a few hundred Mpc for NS-NS mergers as early as 2015. Theoretical
motivation \citep{eichler89,narayan92,rosswog12} and 
observational progress \citep[e.g.][]{gehrels05,barthelmy05a,berger05} 
suggest that at least some short gamma-ray bursts (SGRBs) may be related
to NS-NS mergers.  This hypothesis 
can be proved when both a SGRB and a gravitational wave burst (GWB) are
detected in coincidence with each other in trigger time and direction. 
On the other hand, observations of SGRBs suggest that at least some
of them are collimated \citep[e.g.][]{burrows06,depasquale10}.
Since the strength of the gravitational wave signals does not sensitively 
depend on the orientation of the NS-NS merger orbital plane with respect
to the line of sight,
most GWBs would not be associated with SGRBs even if the SGRB-GWB
association is established. Searching for electromagnetic counterparts
of SGRB-less GWBs is essential to confirm the astrophysical origin
of the GWBs, and to advance our understanding of the compact star merger physics. 
In the literature, an optical ``macronova'' \citep{lipaczynski98,kulkarni05,metzger10}
due to decay of the ejecta launched during the merger\footnote{\cite{kyutoku12}
conjectured that a tip of such an ejecta can reach relativistic speed and give
broad-band afterglow in a wide solid angle.}
and a long-lasting radio afterglow due to interaction between the
ejecta and the ambient medium \citep{nakar11,metzger12,piran12} have
been predicted. Both are challenging to detect \citep{metzger12}. Here we
suggest another possible electromagnetic counterpart of GWBs. We argue
that if the post-merger product is a
short-lived massive neutron star rather than a black hole, a SGRB-less GWB
could be followed by an early X-ray and optical afterglow extending for 
thousands of seconds. We provide observational hints of such a possibility 
in \S2. In \S3, we estimate the duration and brightness of the X-ray and
optical afterglows, and discuss their detectability. 
A brief summary is given in \S4.

\section{Massive neutron star as the post-merger object}

There are two lines of reasoning to suspect that NS-NS mergers can
produce a massive NS rather than a black hole, which may survive for
an extended period of time of the observational interest. The first is
along the line of the observations of NSs and NS-NS binaries in the 
Galaxy (e.g. \cite{lattimer12} for a review).
A secure lower limit of the maximum NS mass is set by
PSR J1614-2230 (in a NS-WD binary) to $1.97\pm 0.04 M_\odot$ 
through a precise measurement of the Shapiro delay \citep{demorest10}. 
NSs with possibly even higher masses, albeit with large uncertainties, 
are also suggested. For example, the NS candidate in the X-ray binary 
4U 1700-377 has a mass $2.44 \pm 0.27 M_\odot$ \citep{rawls11},
and the NS in the NS-WD binary PSR B1516+02B has a mass
$2.08 \pm 0.19 M_\odot$ \citep{freire08}. A stiff equation of state (EOS)
of neutron matter is demanded by the data. Although current data
do not allow to differentiate among various stiff EOS models, most of
these stiff-EOS NS models predict a maximum NS mass close to or higher
than $2.5 M_\odot$ for a non-rotating NS \citep{lattimer12}. 
For rapidly spinning NSs which are likely relevant for the post-merger 
products, the maximum mass can be even higher due to a centrifugal 
support. On the other hand, the observations of the 
Galactic NS-NS systems suggest that the NS mass in these systems
peak at $1.35 M_\odot$, and the sum of the two NS masses for 
a significant fraction of the population is around $2.6 M_\odot$
\citep{lattimer12}. Numerical simulations suggest that 
NS-NS mergers typically eject several percent solar masses \citep{rosswog12}.
As a result, the post-merger products of at lease a fraction
(e.g. $f_{\rm NS} \sim 0.5$) of NS-NS merger
events should have a total mass below the maximum NS mass of a 
rapidly spinning NS. This NS would not collapse until loosing a
significant amount of angular momentum within the characteristic
spin-down time scale. Such a possibility was suggested by
\cite{dai06} and \cite{gao06} to interpret X-ray flares and
plateaus following SGRBs, and is now strengthened by additional data.

The second line of reasoning is based on the observations of the 
SGRB X-ray afterglows. The most direct evidence of a spinning-down
object at the SGRB central engine is in GRB 090515 detected by
Swift \citep{rowlinson10}. After a short
prompt emission phase lasting for $T_{90} = 0.036 \pm 0.016$ s, 
the burst showed an X-ray plateau that lasted for $\sim 240$ s,
after which the flux declines rapidly, and became undetectable by
XRT at $\sim 500$ s after the trigger \citep{rowlinson10}. Such a
steady plateau with rapid decline would be a signature of a 
magnetar at the central engine \citep{zhangmeszaros01a,troja07}.
Even though no redshift measurement was made for this burst, an
analysis suggests that the presumed heavy NS has parameters 
consistent with a magnetar for a reasonable redshift range
\citep{rowlinson10}. A later systematic analysis of Swift SGRB
X-ray lightcurves suggests that a significant fraction of SGRBs
have evidence of an X-ray plateau followed by a steep drop in
flux, which is consistent with a magnetar central engine 
\citep{rowlinson12}. If SGRBs are associated with NS-NS mergers,
it is likely that a millisecond magnetar survived in these SGRBs.

Another indirect piece of evidence is X-ray flares following 
some SGRBs \citep{barthelmy05a}. A possible interpretation is
the magnetic activity of a differentially rotating massive
NS after a NS-NS merger \citep{dai06}. If the magnetic field
strength of this post-merger massive NS is not too high (similar
to that of normal pulsars), the magnetic 
activity of the NS has the right time scale and luminosity to
account for X-ray flares.

\section{Early X-ray and optical afterglow of NS-NS merger-induced GWBs}

At least some SGRBs are collimated \citep{burrows06,depasquale10}.
For the standard X-ray afterglow component (that originates from 
the external shock of the SGRB jet), the afterglow jet opening angle
is believed to be comparable to the prompt emission jet opening
angle, so that a GWB without a SGRB association would have a very faint
``orphan'' afterglow peaking at a time when the jet is decelerated
enough so that the $1/\Gamma$ cone enters line-of-sight.
The prospects of detecting such a SGRB orphan afterglow are poor.
Here we suggest that the afterglow powered by a rapidly spinning
massive NS has a much wider solid angle than the solid angle of the
SGRB jet, so that {\em SGRB-less GWBs can also have a bright  afterglow
from a dissipating proto-magnetar wind with a large solid angle}. 
At the base of the central engine (light cylinder), the wind
launched from the millisecond magnetar is
essentially isotropic. Numerical simulations suggest that this
proto-magnetar wind from a NS-NS merger progenitor would be
collimated by the ejecta launched during the merger process,
but with a much larger angle, $30^\circ-40^\circ$, than the case 
of a massive-star core-collapse progenitor \citep{bucciantini12}.  
This is much larger than the jet opening angle inferred from the 
afterglow modeling of some SGRBs \citep{burrows06,depasquale10}. 
A wider solid angle of proto-magnetar wind than the GRB jet angle 
was also inferred from an analysis of the magnetar engine candidates 
for long GRBs \citep{lyons10}. 

In the following, we adopt the {\em ansatz} that some NS-NS mergers 
produce a massive magnetar. The proto-magnetar wind is
essentially isotropic at the base, with a wide solid angle 
$\theta_{w,1} \sim 40^{\circ}$ for a free wind 
(with a beaming factor $f_{b,w,1}=\Delta \Omega_{w,1} / 4\pi \sim 0.2$) 
and an even larger solid angle $\Delta \Omega_{w,2}$ 
in the equatorial direction for a 
confined wind that pushes the heavy ejecta launched during the merger phase
(with a beaming factor $f_{b,w,2}=\Delta \Omega_{w,2} / 4\pi 
\sim 0.8$, so that the total beaming factor is 
$f_{b,w}=f_{b,w,1}+f_{b,w,2} \sim 1$). This hypothesis
applies regardless of whether the GWB is associated with a SGRB.
If there is a GWB/SGRB association, we expect that SGRB jets 
have a much smaller solid angle. For example, if the typical SGRB jet
opening angle is $\theta_{j} \sim 10^\circ$, one has the jet beaming 
factor $f_{b,j} = \Delta \Omega_j/4\pi \sim 0.015$, so that
$\Delta \Omega_j \ll \Delta \Omega_{w,1} < \Delta \Omega_{w,2}$.

The NS-NS merger event rate is very uncertain. The rate inferred from
the Galactic NS-NS systems has a wide range $2 - 2\times 10^{4} 
~{\rm Gpc^{-3}~yr^{-1}}$ \citep{phinney91,kalogera04,abadie10}. This is
consistent with the upper limit $2\times 10^5~{\rm Gpc^{-3}
~yr^{-1}}$ set by the current non-detection with the last LIGO and
VIRGO run \citep{ligo}. Within the advanced LIGO horizon $\sim 300$ 
Mpc, the NS-NS merger rate (and therefore GWB rate) would be 
$R_{\rm GWB}\sim (0.2 - 2000) ~{\rm yr^{-1}}$. Among these, 
$R_{\rm GWB-ag}\sim (0.1 - 1000) (f_{\rm NS}/0.5) (f_{b,w}) 
~{\rm yr^{-1}}$ would have strong afterglow emission associated
with the proto-magnetar wind, most of which would not have a SGRB 
association, since the line of sight is outside the SGRB cone even
if there is a SGRB/GWB association.

After the merger, the proto-NS is initially very hot and cools via
neutrino emission. After about 10 seconds, the NS is cooled enough
so that a Poynting-flux-dominated outflow can be launched
\citep{usov92,metzger11}. It will be spun down by magnetic dipole radiation 
and by the torque of a strong electron-positron pair wind 
flowing out from the magnetosphere. 
Since before the merger the two NSs are in the Keplerian orbits, the
post-merger product should be near the break-up limit. 
We take $P_0=1~{\rm ms}~P_{0,-3}$ as a typical value
of the initial spin period of the proto-magnetar. An uncertain parameter is 
the polar-cap magnetic field of the dipole magnetic field component, 
$B_{p}$,  which depends on whether the $\alpha-\Omega$
dynamo is efficiently operating, and on the magnetic field strength of 
the parent NSs if the dynamo mechanism is not efficient. Given
nearly the same amount of the total rotation energy 
$E_{\rm rot} = (1/2) I \Omega_0^2 \sim 2\times 10^{52}~{\rm erg}$
($\Omega_0 = 2\pi/P$),
the luminosity, and hence, the afterglow flux critically depend on 
$B_p$. As a rough estimate, we apply the dipole spindown formula.
Correcting for the beaming factor $f_w$ and the 
efficiency factor $\eta_x$ to convert the spin
down luminosity to the observed X-ray luminosity in the detector
band, one gets
\begin{eqnarray}
F_x & = & \frac{\eta_x L_{sd}}{4 \pi f_{b,w} D_L^2}
\simeq 2\times 10^{-8}~{\rm erg~s^{-1}~cm^{-2}} \nonumber \\
 & \times & \eta_{x,-2} f_{b,w}^{-1} \left(\frac{D_L}{300~{\rm Mpc}}
\right)^{-2} I_{45} P_{0,3}^{-2} T_{sd,3}^{-1},
\label{Fx}
\end{eqnarray}
where $L_{sd} = I \Omega_0^2 / (2 T_{sd})$ is the characteristic
spindown luminosity, and 
\begin{equation}
 T_{sd} \simeq 2\times 10^3~{\rm s}~ I_{45} B_{p,15}^{-2} P_{0,-3}^2
R_6^{-6}
\label{Tsd}
\end{equation}
is the characteristic spindown time scale. Here $I = 10^{45} I_{45}$ 
is the moment of inertia (typical value $I_{45} = 1.5$ for a massive 
NS), $R = 10^6 R_6$ is the radius of the NS, and the convention
$Q_x = Q/10^x$ has been adopted. Here we have assumed that a good
fraction ($\eta_x \sim 0.01$) of spin down energy is released in the 
X-ray band. This is based on the 
following two considerations: First, some SGRBs indeed have a bright 
X-ray plateau that is likely due to the magnetar spindown origin 
\citep{rowlinson10,rowlinson12}, which suggests that the main
energy channel of releasing the magnetic dissipation energy is
in the X-ray band. Second, a rough theoretical estimate shows that
the typical energy band of a dissipating magnetized wind with a
photon luminosity $L_\gamma \sim 10^{49}~{\rm erg~s^{-1}}$ could be 
in X-rays. 

We consider two mechanisms to dissipate the magnetar wind energy to
radiation. (1) In the free wind zone with solid angle $\Delta \Omega_{w,1}$,
one may consider a magnetized wind with 
a luminosity $L_w$ and magnetization
parameter $\sigma(R)$ dissipated at a radius $R$ from the central
engine. Assuming that the magnetic energy is abruptly converted to
the internal energy of power-law distributed electrons (such as in
the scenario of the ICMART model), one can generally estimate the
typical synchrotron energy as \citep{zhangyan11}
$E_p \simeq 80~{\rm keV}~ L_{\gamma,48}^{1/2} R_{15}^{-1} \eta_{x,-2}^{3/2}
\sigma_4^2$. A cooled-down proto-magnetar typically has $\sigma_0
\sim 10^9$ at the central engine \citep{metzger11}. A magnetized
flow can be quickly accelerated to $\Gamma \sim \sigma_0^{1/3}
\sim 10^3$ at $R_0 \sim 10^7$ cm, where $\sigma \sim \sigma_0^{2/3}
\sim 10^6$ \citep{komissarov09}.  After this phase, the flow may
still accelerate as $\Gamma \propto R^{1/3}$, with $\sigma$ falling
as $\propto R^{-1/3}$ \citep{drenkhahn02}. At $R \sim 10^{15}$ cm,
one has $\sigma \sim 2\times 10^3$, so that $E_p \sim 3.7$ keV,
which is in the X-ray band.
(2) One can also consider the confined magnetar wind zone with
solid angle $\Delta \Omega_{w,2}$ where the magnetar wind is
expanding into a heavy ejecta launched during the merger 
process\footnote{I thank Xue-Feng Wu for pointing out this
possibility.}. The magnetic energy may be rapidly discharged
upon interaction between the wind and the ejecta, which occurs
at a radius $R \sim v t_{\rm delay} = 3\times 10^{10}~{\rm cm}
(v/0.1 c) t_{\rm delay,1}$, where $v \sim 0.1 c$ is the speed of
ejecta, and $t_{\rm delay} \sim 10$ s is the delay time between
the merger and the launch of a high-$\sigma$ magnetar wind. 
The Thomson optical 
depth for a photon to pass through the ejecta shell is $\tau_{th} \sim
\sigma_{\rm T} M_{\rm ej} / (4\pi R^2 m_p) \sim 7\times 10^8
(M_{ej}/(0.01M_\odot)) \gg 1$.
So the spectrum of the dissipated wind is thermal-like. One can
estimate the typical energy $\sim k (L_w/4\pi R^2 \sigma)^{1/4}
\sim 5~{\rm keV} L_{w,49}^{1/4} (R/(3\times 10^{10}~{\rm cm}))^{-1/2}$,
which is also in the X-ray band.

One can see that the X-ray band flux of the early afterglow
(Eq.\ref{Fx}) is very high, well above the sensitivity threshold
of Swift XRT \citep{moretti07}
\begin{equation}
 F_{x,th} = 6 \times 10^{-12}~{\rm erg~s^{-1}~cm^{-2}}~ T_{\rm obs,2}^{-1},
\end{equation}
where $T_{\rm obs}$ is the observation time. 
 The light curve is expected to be flat (a plateau)
lasting for a duration $T_{sd}$ followed by a $t^{-2}$ decay.
However, since the NS spins down quickly in the $t^{-2}$ regime,
it is likely that it would lose centrifugal support and collapse
to a black hole shortly after the end of the plateau. In this
case, essentially all the materials collapse into the BH, without
substantial accretion afterwards. The light curve then shows a
very sharp drop in flux at the end of the plateau, 
similar to what is seen in GRB 090515 \citep{rowlinson10}.

The challenge to detect such a bright X-ray afterglow following
a GWB is its short duration (Eq.\ref{Tsd}) and the large error box
of a GWB trigger. This requires a Swift-like space detector 
for quick slew, but the error box of the
GWB trigger, typically a few tens to a hundred 
square degrees \citep{abadie12},
is much larger than the XRT field of view 
(0.16 square degree). How to efficiently search for the bright
X-ray source within $T_{sd}$ in such a large sky area is 
challenging. Even though some strategies using Swift have been proposed
\citep{kanner12}, the current searches for the GWB afterglow
typically happen about half-day after the GWB trigger \citep{evans12}.
The problem can be alleviated if $B_p$ of the proto-NS is weaker. 
For example, even for a typical pulsar field $B_p \sim 10^{12}$ G 
\citep{dailu98b,dai06}, the X-ray luminosity can be still
as high as $5\times 10^{-11} ~{\rm erg~s^{-1}~cm^{-2}}$ for detection,
while the duration of the plateau extends to $T_{sd} \sim 2
\times 10^7$ s. This would give enough time to search for the
X-ray afterglow. However, strong magnetic
fields are likely generated during the merger events \citep{price06}.
Very likely one has to face the large-error-box, short-duration
problem. An ideal strategy to observe this early afterglow is to 
design a large field-of-view imaging X-ray telescope, preferably 
with fast-slewing capability. Such a telescope, even 
with a moderate sensitivity, can catch the bright
early X-ray afterglows of SGRB-less GWBs. The new mission concept
ISS-Lobster \citep{gehrels12} invokes an X-ray wide-field imager
with a 0.5-sr field of view that covers $\sim 50\%$ of the sky
every 3 hours, which is ideal to detect this bright X-ray afterglow.

The optical flux of the proto-magnetar wind is subject to 
uncertainties. In the free wind zone (solid angle $\Delta \Omega_{w,1}$),
the emission spectral shape is synchrotron.  If one has the standard $F_\nu 
\propto \nu^{1/3}$ synchrotron spectrum below $E_p$, the specific
X-ray flux at 1 keV $F_\nu({\rm X}) \sim 4$ mJy would correspond to
a R-band magnitude 17. This would be an optimistic estimate of
the optical brightness. For the confined wind zone (solid angle
$\Delta \Omega_{w,2}$), the spectrum of a dissipating wind
is quasi-thermal, and the optical flux is greatly suppressed.
Indeed, no bright optical emission was detected during the plateau
phase of GRB 090515 \citep{rowlinson10}, suggesting that the
optical emission of a dissipative proto-magnetar wind is
suppressed. Nonetheless, the interaction between the magnetar 
wind and the ejecta in the confined wind zone can give very interesting
radiation signatures in the optical band (H. Gao et al. 2013, in
preparation). Wide-field optical telescopes are essential to search 
for such optical GWB afterglows in the large GWB error box. 

The gravitational wave signals from these GWBs have an interesting 
signature: after the standard chirp signal during the in-spiral and 
merger phases \citep{flanagan98,kobayashimeszaros03}, there should be 
an extended GW emission episode afterwards due to a secular bar-mode 
instability of the newly formed proto-magnetar 
\citep{corsi09}. The signature is in the advanced LIGO 
frequency band, and can in principle be detected. Jointly 
detecting such a GW signal along with the X-ray afterglow
would give an unambiguous identification of the proto-magnetar
nature of the central engine. 

Some SGRBs are followed by an extended emission, which sometimes can be
very bright \citep[e.g.][]{gehrels06}. It is unclear whether the extended
emission shares the same solid angle with the short hard spikes.
If it has a wider solid angle than the short hard spike emission, 
as expected in the magnetar engine scenario 
\citep[e.g.][]{metzger08}, then such a bright extended emission
(lasting $\sim 100$ s) can be also associated with SGRB-less GWBs.
This emission is brighter than the X-ray afterglow emission discussed 
above, and can be readily detected by wide-field imagers such as 
ISS-Lobster.

\section{Summary}

We have proposed another electromagnetic counterpart of GWBs from
NS-NS mergers. It applies to the cases when the two
NSs are not very massive (as observed in Galactic double NS systems), 
so that the post-merger product has a mass
below the maximum mass of a rapidly spinning NS. We show that such
a scenario is plausible in view of the observations of Galactic
NSs, NS-NS systems, and SGRB afterglows. The proto-magnetar
would eject a wide-beam wind, whose dissipation
would power an X-ray afterglow as bright as $\sim (10^{-8}-10^{-7})
~{\rm erg~s^{-1}~cm^{-2}}$. The duration is typically $10^3 - 10^4$ s, 
depending on the strength of the dipolar magnetic fields. It is
challenging to detect the X-ray afterglow with the current facilities 
such as Swift, but a wide-field X-ray imager (such as ISS-Lobster) 
would be ideal to catch this bright X-ray
signal. The optical afterglow flux is subject to large uncertainties,
but could be as bright as 17th magnitude in R band. Prompt, deep
optical follow-up observations of GWBs are desirable.
The detection of these signals would confirm
the astrophysical origin of GWBs, and shed light into the physics
of NS-NS mergers and the NS equation of state.




\acknowledgments
I thank a Cheung Kong scholarship of China, the hospitality of the 
KIAA and Department of Astronomy of Peking University, and the 
sabbatical committee of the UNLV faculty senate, to provide me an 
ideal working environment to conduct research efficiently.  
I thank stimulative discussion with Xue-Feng Wu, He Gao, Zi-Gao Dai,
and Yi-Zhong Fan, and helpful comments from Kunihito Ioka, Elenora
Troja and an anonymous referee. This work is partially supported by
NSF AST-0908362.

\end{document}